\documentstyle[12pt]{article}

\begin{document}
\begin{titlepage}
\title{\Large\bf Black Hole with Non-Commutative Hair}
\vskip 3 cm
\author{~\\~\\C.~Klim\v{c}\'{\i}k$^1$, P.~Koln\'{\i}k$^1$, and
A.~Pompo\v{s}$^2$\\
{}~\\~\\ \small $^1$ Theory Division of the Nuclear Centre,
Charles University,\\
\small  V Hole\v{s}ovi\v{c}k\'ach 2, 180 00
Prague 8, Czech Republic\\~\\
\small $^2$ Department of Theoretical Physics, Charles University,\\
\small V Hole\v{s}ovi\v{c}k\'ach 2, 180 00
Prague 8, Czech Republic\\~\\}
\date{\small June, 1993}
\maketitle
\begin{abstract}
The specific nonlinear vector $\sigma$-model coupled to Einstein
gravity
is investigated. The model arises in the studies of the gravitating
matter in non-commutative geometry. The static spherically symmetric
spacetimes are identified by direct solving of the field equations.
The asymptotically flat black hole with the ``non-commutative''
vector hair appears for the
special choice of the integration constants, giving thus another
counterexample to the famous ``no-hair'' theorem.
\end{abstract}
\centerline{PACS index: 04.20, 04.30}
\end{titlepage}


\section{Introduction}

The black-hole solutions of the Einstein equations have constituted
one of
the most fascinating feature, with which Einstein's general
relativity
has enriched the theoretical physics. The curvature
singularities and the event horizons  have been qualitatively novel
structures, the role of which in our world is still to be clarified,
in particular in quantum context. Moreover, the black holes are in a
sense exclusive, as the famous ``no-hair" theorem says.
The developments in eighties led to considerable interest in the
various
``exotic"  gravitating matter theories. For example,
the important role of the classical solutions of the pure Yang-Mills
theory had to be elucidated from the point of view of the coupling
to gravity and the string theory effective action also incorporates
the matter described by the higher spin fields.
As the product of those
efforts, the possibility of the higher spin hair in the
Einstein-Yang-Mills,
Einstein-Skyrme and Einstein-axion equations was discovered
\cite{bis,GHS,CKO}.

In this contribution, we give another ``counterexample"  to the
no-hair
theorem, showing explicitly the black hole supporting the
``non-commutative'' hair in 3+1 dimensions. The action of the system
is that of the nonlinear
vector $\sigma$-model coupled to the Einstein gravity. The explicit
form
of the action is inspired, in fact, by the recent studies of the
gravitating matter in the non-commutative geometry. The
non-commutative
pure Einstein-Hilbert action for the Connes' double-sheeted
manifolds
gives the specific nonlinear vector $\sigma$-model coupled to the
Einstein
gravity.
This theory then possesses the black hole
solution with the $\sigma$-model vector field hair.

In what follows, we present the detailed way of solving of the
field equations
for the static spherically symmetric ansatz. We pick up the black
hole
solution and provide a detailed analysis of the curvature tensor and
regularity condition for the hair at the asymptotic region, near the
horizon and below the horizon.

\section{The Model and the Solutions of the Field Equations}

We wish to find static spherically symmetric
solutions of the gravitating nonlinear vector $\sigma$-model
with the (non-commutative Einstein-Hilbert) action \cite{KPS}
\begin{equation}
I=\int_Y d^4x\sqrt{-g}\left[2R+Q^{\alpha\beta\gamma\delta}(V)
D_{\alpha}V_{\beta}D_{\gamma}V_{\delta}\right],
\end{equation}
where
\begin{equation}
Q^{\alpha\beta\gamma\delta}(V)=\frac{4}{(V^2)^3}
(V^\alpha V^\beta V^\gamma V^\delta - g^{\alpha\beta}V^\gamma
V^\delta V^2 -g^{\gamma\beta} V^\alpha V^\delta V^2),
\end{equation}
and $D_\alpha$ is the covariant derivative.
This vector $\sigma$-model action can be rewritten as
\begin{equation}
I=\int_Y d^4x\sqrt{-g}\left[2R+4\kappa D_\beta(\frac{V^\alpha
V^\beta}
{\sqrt{V^2}})D_\alpha (\frac{1}{\sqrt{V^2}})\right],\label{old}
\end{equation}
with $\kappa=1$.
In what follows, we shall consider a general gravitational
constant $\kappa>0$.
Let us introduce the Schwarzschild-like coordinates, in
which the metric has a form
$$
ds^2=-e^{\nu(r)}dt^2+e^{\lambda(r)}dr^2+r^2(d\theta^2+
\sin^2\theta d\varphi^2).
$$
We shall look for the solutions with $V^{0}$ and $V^1$ as
the only non-zero components (0 and 1 correspond to $t$ and $r$
respectively). Assume also that $V^2$ is
positive.
Then $V^{\alpha}$ can be renormalized, introducing new functions
$f^{\alpha}(r)$, $\sigma(r)$
$$f^{\alpha}=\frac{V^{\alpha}}{\sqrt{V^2}}~,~~\sigma=\frac{1}
{\sqrt{V^2}}.$$
Now, the action can be rewritten, using a Lagrange multiplier
$\Lambda$
$$
I=\int_Y d^4x \sqrt{-g}\left[2R-4\kappa f^{\alpha}f^{\beta}\frac{D_
{\alpha}
D_{\beta}\sigma}{\sigma}-4\kappa \Lambda(f^{\alpha}f{_\alpha}-1)
\right].
$$
Variation of this action with respect to all variables yields the
equations of motion for this model
\begin{eqnarray}
f^{\alpha}f^{\beta}g_{\alpha\beta}&=&1,\nonumber\\
f^{\alpha}\left(\frac{D_{\mu}D_{\alpha}\sigma}{\sigma}+ \Lambda
g_{\mu\alpha}\right)&=&0,\nonumber\\
D_{\beta}\left(2f^{\alpha}f^{\beta}\frac{D_{\alpha}\sigma}{\sigma}-
D_{\alpha}
(f^{\alpha}f^{\beta})\right)&=&0,\label{2.6}
\end{eqnarray}
\begin{eqnarray}
R_{\mu\nu}-{\textstyle\frac12} g_{\mu\nu}R&=&\kappa\Bigg[-g_{\mu\nu}
\left(f^{\alpha}f^{\beta}\left(\frac{D_{\alpha}D_{\beta}\sigma}
{\sigma}+
\Lambda g_{\alpha\beta}\right)-\Lambda\right)+
\nonumber\\
& &+2f^{\alpha}f_{~[\mu}\left(\frac{D_{\alpha}
D_{\nu]}\sigma}{\sigma}
+\frac{D_{\alpha}\sigma D_{\nu]}\sigma}{\sigma^2}
\right)+2f_\mu f_\nu\Lambda+
\nonumber\\
& &+D^{\alpha}
\left(f_{\mu}f_{\nu}
\frac{D_{\alpha}\sigma}{\sigma}\right)-2
D_{\alpha}(f^{\alpha}f_{~[\nu})
\frac{D_{\mu]}\sigma}{\sigma}\Bigg].\nonumber
\end{eqnarray}
where $[\alpha~\beta]$ means the symmetrization in the
indices.
In our ansatz these general equations acquire the following
form (the equation (\ref{2.11}) is in fact the first integral of
(\ref{2.6}))
\begin{eqnarray}
-e^\nu f^0f^{0}+e^\lambda f^1f^{1}&=&1,\label{vf}\\
\frac{\sigma'}{\sigma}\frac{\nu'}{2}e^{-\lambda}+\Lambda&=&0,
\label{2.9}\\
\frac 1 \sigma \left(\sigma''-\frac{\lambda'}{2}\sigma'\right)+
e^\lambda\Lambda&=&0,\label{2.10}
\end{eqnarray}
\begin{equation}
{r^2} e^{\frac{\nu+\lambda}{2}}\left[
\left(f^1f^1\right)'+f^1f^1\left(\frac{\nu'}2+\lambda'-
\frac{2\sigma'}{\sigma}
+\frac 2 r \right)+f^0f^0\frac{\nu'}2 e^{\nu-\lambda}\right]=A
,\label{2.11}
\end{equation}
\begin{eqnarray}
\frac 1 {r^2}-e^{-\lambda}\left(\frac 1 {r^2}-\frac{\lambda'}{r}
\right) &=&\kappa\Bigg[f^0f^0\left(-\frac{\nu'}{2}
e^{\nu-\lambda}\frac{\sigma'}{\sigma}-\Lambda e^\nu\right)+
\nonumber\\
& &~~~~+f^1f^1\left(\frac{\sigma''}{\sigma}-\frac{\lambda'}{2}
\frac{\sigma'}{\sigma}+\Lambda e^\lambda \right)-\Lambda\Bigg]+
\nonumber\\
& &+\kappa \left(f^0f^0\right)'\frac{\sigma'}{\sigma}e^{\nu-\lambda}
+\nonumber\\
& &+\kappa f^0f^0e^{\nu-\lambda}\Bigg[\frac{\sigma'}{\sigma}
\left(\frac 5 2 \nu'-\frac{\sigma'}{\sigma}-\frac{\lambda'}{2}
+\frac 2 r \right)+\nonumber\\
& &~~~~~~~~~~~+\frac{\sigma''}{\sigma}+2\Lambda
e^\lambda\Bigg],\label{2.12}
\end{eqnarray}
\begin{eqnarray}
\frac 1 {r^2} -e^{-\lambda}\left(\frac{1}{r^2}+\frac{\nu'}{r}
\right)&=&
\kappa \Bigg[f^0f^0\left(-\frac{\nu'}{2}
e^{\nu-\lambda}\frac{\sigma'}{\sigma}-\Lambda e^\nu\right)+
\nonumber\\
& &~~~~~~~+f^1f^1\left(\frac{\sigma''}{\sigma}-\frac{\lambda'}{2}
\frac{\sigma'}{\sigma}+\Lambda e^\lambda \right)-\Lambda\Bigg]+
\nonumber\\
& &+\kappa\left(f^1f^1\right)'\frac{\sigma'}{\sigma}+
\kappa f^0f^0\nu'e^{\nu-\lambda}\frac{\sigma'}{\sigma}+
\nonumber\\
& &+\kappa f^1f^1\Bigg[\frac{\sigma'}{\sigma}\left
(-\frac{\sigma'}{\sigma}
+\frac{5}{2} \lambda'+
\frac{\nu'}{2}+\frac{2}{r}
\right)-\nonumber\\
& &~~~~~~~~~~-3\frac{\sigma''}{\sigma}-
2e^\lambda\Lambda\Bigg], \label{2.13}
\end{eqnarray}
\begin{eqnarray}
e^{-\lambda}\left(\nu''+\frac{(\nu')^2}{2} +\frac{\nu'-\lambda'} r
-\frac{\nu'\lambda'}{2}\right)=2\kappa\Lambda,\label{2.14}
\end{eqnarray}
where $A$ in Eq.~(\ref{2.11}) is an integration constant.

These equations look formidable, nevertheless, they can be
patiently disentangled as follows.
{}From Eq.~(\ref{2.9}) and Eq.~(\ref{2.10}) we can write
$$2\sigma''=\sigma'(\nu'+\lambda'),$$
thus (if $\sigma'\ne 0$)\footnote{In the case $\sigma'=0$ we end up
with the standard Schwarzschild metric.}
\begin{equation}
e^\nu =\sigma'^2 e^{-\lambda} e^{-2K},\label{ni}
\end{equation}
where $K$ is a constant.
Substituting Eq.~(\ref{ni}) and Eq.~(\ref{vf}) into Eq.~(\ref{2.11})
we obtain
\begin{equation}
\left(f^1f^1\right)'+2f^1f^1\left(\frac{1}r+\frac{\sigma''}{\sigma'}
-\frac{\sigma'}\sigma\right)=\frac{A\,e^K}{r^2\sigma'}+e^{-\lambda}
\left(\frac{\sigma''}{\sigma'}-\frac{\lambda'}2\right).\label{f12'}
\end{equation}
The difference of Eq.~(\ref{2.12}) and Eq.~(\ref{2.13})
results in
\begin{equation}
\frac{\sigma''}\sigma +\kappa\left(\frac{\sigma'^2}{\sigma^2}-
\frac{r}{2}
\frac{\sigma'^3}{\sigma^3}\right)=0,\label{sigma}\end{equation}
and Eq.~(\ref{2.13}), combined with Eq.~(\ref{f12'}), yields
\begin{equation} f^1f^1=\frac{\sigma^2}{\kappa r^2\sigma'^2}-
\frac{e^{-\lambda}}{\kappa}\left(\frac
{\sigma^2}{r^2\sigma'^2}+\frac{2\sigma^2\sigma''}{r\sigma'^3}-\frac
{\lambda'\sigma^2}{r\sigma'^2}\right)-\frac{A\sigma e^K}{\sigma'^2
r^2}.\label{f12}\end{equation}
Using Eq.~(\ref{vf}) and inserting $f^1f^1$ from Eq.~(\ref{f12}) into
Eq.~(\ref{2.6}), we obtain
\begin{equation}
\frac{\sigma^2}{\sigma'^2}Q''+\left(\kappa\frac\sigma{\sigma'}-
\frac{\sigma''\sigma^2}{\sigma'^3}\right)Q'-
\kappa\frac{Q}{2}=0,\label{kve}\end{equation}
where we have introduced $Q(r)$, defined by
$$Q(r)\equiv e^{\nu(r)}r.$$
Both equations (\ref{sigma}) and (\ref{kve}) can be easily solved
after changing variable $r\to\sigma$, as {\it then} they become
\begin{eqnarray}
\frac{d^2 r}{d\sigma^2}-\frac{\kappa}\sigma\frac{dr}{d\sigma}+
\frac{\kappa r}
{2\sigma^2}&=&0,\label{forr}\\
\frac{d^2 Q}{d\sigma^2}+\frac{\kappa}\sigma\frac{dQ}{d\sigma}-
\frac{\kappa Q}
{2\sigma^2}&=&0.\label{forQ}
\end{eqnarray}
The general solution of Eq.~(\ref{forr}) and Eq.~(\ref{forQ}) is
\begin{eqnarray}
r&=&c_1\sigma^{\alpha_1}+c_2\sigma^{\alpha_2},\label{r}\\
Q&=&k_1\sigma^{\beta_1}+k_2\sigma^{\beta_2},\label{Q}
\end{eqnarray}
where $\alpha_{1,2}=(1+\kappa\pm \sqrt{\kappa^2+1})/2 $,
 $\beta_{1,2}=(1-\kappa\pm \sqrt{\kappa^2+1})/2$,
and $c_{1,2}$ and $k_{1,2}$ are (real) constants.
Formulae (\ref{ni}), (\ref{r}), and (\ref{Q}) fully determine
the metric, and the hair $f^1f^1$ is given by Eq.~(\ref{f12})
and $f^0f^0$ by Eq.~(\ref{vf}). It it is easy to check that this
solution solves also Eq.~(\ref{2.14}),
hence, it is the most general solution for our
ansatz.
Let us study the behavior of the scalar curvature of the
solutions. We have
$$R=\frac{2}{r^2}-e^{-\lambda}\left(\frac{2}{r^2}+
\frac{\nu'^2}{2}+\nu''+2\frac{\nu'-\lambda'}{r}-
\frac{\nu'\lambda'}{2}\right).$$
This expression diverges for $r\to 0$ and tends to zero
for $r\to\infty$.
Is there any special choice of parameters giving the black-hole
solutions?
The answer is affirmative. Indeed, set $c_1=0$, $c_2>0$, $k_1>0$ and
$k_2<0$,
then from Eq.~(\ref{r}) it can be explicitly extracted $\sigma(r)$
and Eq.~(\ref{Q}) and Eq.~(\ref{ni}) yield
\begin{eqnarray}
e^\nu=\frac{k_1}{r}\left(\frac{r}{c_2}\right)^{\frac{1+
\sqrt{\kappa^2+1}}{\kappa}} +
\frac{k_2}{r}\left(\frac{r}{c_2}\right)^{-\kappa-\sqrt{
\kappa^2+1}}\label{en},
\end{eqnarray}
\begin{eqnarray}
e^\lambda=\frac{e^{-2K}}{\alpha_2^2 r}\frac 1 {k_1\left(
\frac{r}{c_2}
\right)^{-\frac{2\kappa+1+\sqrt{\kappa^2+1}}{\kappa}}+k_2\left(
\frac r{c_2}\right)^{-\frac{\kappa^2+2\kappa+2+(2+\kappa)\sqrt{
\kappa^2+1}}{\kappa}}}~.\label{el}
\end{eqnarray}
There is obvious horizon for
$$
r=c_2\left(-\frac{k_2}{k_1}\right)^\frac{\kappa}{\kappa^2+1+
(1+\kappa)
\sqrt{\kappa^2+1}}.
$$
{}From the formulae (\ref{en}) and (\ref{el}) it turns
out that the scalar curvature is bounded at the horizon.
Let us analyze the behavior of the curvature tensor in detail.
Following
orthonormal vierbein is parallelly propagated along the geodesics
respecting spherical symmetry
\begin{eqnarray}
e^{(0)}_\mu&=&(e^{\nu/2},0,0,0),\nonumber\\
e^{(1)}_\mu&=&(0,e^{\lambda/2},0,0),\nonumber\\
e^{(2)}_\mu&=&(0,0,r,0),\nonumber\\
e^{(3)}_\mu&=&(0,0,0,r \sin \theta).\nonumber
\end{eqnarray}
The explicit evaluation of all (nontrivial) vierbein components
of Riemann tensor gives the following result. All of them
vanish for
$r\to\infty$,
are finite at horizon, and diverge for $r\to 0$.
We conclude that the metric (\ref{en}), (\ref{el}) is black hole
metric for $c_2>0$, $k_1>0$ and $k_2<0$.
It is asymptotically flat, singular at $r=0$ and regular at the
horizon.
It remains to analyze the behavior of the remaining component
$V^\alpha$ of the non-commutative metric.

First of all, we discuss the behavior of $V^2$, which is the only
independent scalar quantity which can be built up from the field
$V^\alpha$. It is very simple
\begin{equation}
V^2=\left(\frac{r}{c_2}\right)^{-\frac{2\kappa+2+2\sqrt{\kappa^2+1}}
{\kappa}}.\label{V}
\end{equation}
Hence $V^2$ (the distance of the Connes' sheets) tends to zero
for $r\to\infty$, it is bounded near the horizon and diverges for
$r\to 0$. Thus the behavior of $V^2$ at infinity and near the
horizon is satisfactory.
As far as the behavior of the components is concerned, from
Eq.~(\ref{V}) it is easy to identify the explicit form of the
$\left(V^1\right)^2$ for the black-hole metric
Eq.~(\ref{en}), (\ref{el}).
It reads
\begin{eqnarray}
\left(V^1\right)^2&=&\frac{\kappa^2+\kappa+1-(1+\kappa)\sqrt
{\kappa^2+1}}{2}
\left(\frac{r}{c_2}\right)^{-\frac{2\kappa+2+2\sqrt{\kappa^2+1}}
{\kappa}}
\times\nonumber\\
& &\times\Bigg[1+c_2k_2\frac{\sqrt{\kappa^2+1}-1}2 e^{2K}
\left(\frac r{c_2}\right)^{-\frac{\kappa^2+\kappa+2+(\kappa+2)
\sqrt{\kappa^2+1}}
{\kappa}}-\nonumber\\
& &-\left(A e^K+c_2k_1
\frac{\sqrt{\kappa^2+1}-1}2 e^{2K}\right)
\left(\frac{r}{c_2}\right)^{-\frac{\kappa+1+\sqrt{\kappa^2+1}}
{\kappa}}\Bigg].\nonumber
\end{eqnarray}
Then $\left(V^0\right)^2$ is given by
\begin{eqnarray}
\left(V^0\right)^2&=&e^{-2K}\left[k_1\left(\frac r{c_2}
\right)^{\frac{1+\sqrt{\kappa^2+1}}{\kappa}}+k_2\left(\frac
r{c_2}\right)^{-\kappa-\sqrt{\kappa^2+1}}
\right]^{-2}\times\nonumber\\
& &\times\Bigg[1+c_2k_2e^{2K}\frac{\sqrt{\kappa^2+1}-3}{2}
\left(\frac r{c_2}
\right)^{-\frac{\kappa^2+\kappa+2+(2+\kappa)\sqrt{\kappa^2+1}}
{\kappa}}-\nonumber\\
& &-\left(c_2k_1e^{2K}\frac{\sqrt{\kappa^2+1}+1}{2}+A e^K
\right)\left(
\frac r{c_2}\right)^{-\frac{\kappa+1+\sqrt{\kappa^2+1}}{\kappa}}
\Bigg].\label{V0}
\end{eqnarray}
Which choice of the integration constants ensures the regularity
of the hair? The non-commutative geometry requires that the
components of the non-commutative metric $g_{\alpha\beta}$,
$V_\alpha$ must be real \cite{KPS,CFF}.
But  $\left(V^1\right)^2$ is positive only for $r>r_{cr}$ where
$r_{cr}$ is given implicitly by the formula
\begin{eqnarray}
A&=&c_2 k_2
\frac{\sqrt{\kappa^2+1}-1}2 e^{K}
\left(\frac{r_{cr}}{c_2}\right)^{-\frac{\kappa^2+1+(\kappa+1)
\sqrt{\kappa^2+1}}{\kappa}}-\nonumber\\
& &-c_2k_1\frac{\sqrt{\kappa^2+1}-1}2 e^{K}+e^{-K}
\left(\frac{r_{cr}}{c_2}\right)^{\frac{\kappa+1+\sqrt
{\kappa^2+1}}{\kappa}}.\label{2.25}
\end{eqnarray}
Therefore, from the geometrical point of view, the
space-time becomes singular at $r=r_{cr}>0$ and not
at $r=0$ as it could seem at the first sight. There is no
curvature singularity at $r_{cr}$, however, the (non-commutative)
metric ceases to be real \cite{KPS,CFF}.
The parameters of the general solution have to be chosen in such way
that $r_{cr}$ is below the horizon, otherwise we would loose the
black hole. Moreover, we have to ensure that $\left(V^0\right)^2$ is
positive
for $r>r_{cr}$. The following choice of the integration constants
meets these criteria
\begin{eqnarray}
\left(-\frac{k_2}{k_1}\right)^{\frac 1{\sqrt{\kappa^2+1}}}
\ge e^{2K} c_2k_1\frac{3\sqrt{\kappa^2+1}-\kappa^2-1}{2},\nonumber\\
r_{cr}(A)<c_2 \left(-\frac{k_2}{k_1}\right)^{\frac{\kappa}
{\kappa^2+1+(\kappa+1)\sqrt{\kappa^2+1}}},
\nonumber
\end{eqnarray}
where the dependence $r_{cr}(A)$ is given by Eq.~(\ref{2.25}).

The next subtle point consists in a behavior of the
$\left(V^0\right)^2$ on the horizon. This quantity is obviously
divergent, as it can be seen from Eq.~(\ref{V0}). Is this divergence
pathological? It need not to be necessarily so, because the
invariant quantity $V^2$ is smooth and bounded. Moreover, as the
experience from the Bekenstein (conformal scalar) black-hole teaches
\cite{bek1}, even the diverging (invariant) hair on the
horizon need
not mean a pathology. Certainly, the careful and independent
analysis of the issue is required as in the Bekenstein case
\cite{bek2}.
However, apart from the regular behavior of $V^2$ on the horizon,
we may give another invariant quantity which is regular on the
horizon, namely, the non-commutative Ricci scalar or---what
is the same thing---the Lagrangian of our model.

\section{Conclusion}

We have solved the field equations of the nonlinear vector
$\sigma$-model (\ref{old}) for the static spherically
symmetric ansatz with $V^\alpha$ spacelike\footnote{
In the time-like case the hair singularities cannot be hidden
beyond the horizon.}.
We picked up the particular black-hole solution which is
asymptotically flat, it has the hair asymptotically falling
off and it has the smooth horizon covering the singularity
inside.

The singularity is not the standard curvature singularity,
but it corresponds to the point where the components of
the non-commutative vielbein (see \cite{KPS}) become
complex. The situation is somewhat analogous to that
reported in \cite{PT}. In that case, there occured a specific
singular behavior between the horizon and the curvature
singularity, where the metric has lost its proper signature.

We hope to understand better the origin of the peculiar
singularities inside the black-hole in our future works.

\end{document}